\documentclass[letterpaper,preprintnumbers,twocolumn,superscriptaddress,aps,nofootinbib]{revtex4}
\usepackage{amssymb}
\usepackage[centertags]{amsmath}
\usepackage{txfonts}
\usepackage{epsfig}
\usepackage{bm}
\usepackage{color}
\usepackage{graphicx,graphics}
\usepackage{multirow}
\usepackage{float}
\usepackage[pdfstartview=FitH]{hyperref}
\hypersetup{colorlinks=true, citecolor=blue, linkcolor=blue,filecolor=black,urlcolor=blue}
\allowdisplaybreaks[2]

\usepackage{slashed}
\usepackage{ulem}

\usepackage{booktabs}
\usepackage{multirow}

\begin{document}

\title{Global Polarization in the Extremely Rapidly Rotating QGP in HIC}

\author{Zuo-Tang Liang}
\affiliation{Key Laboratory of Particle Physics and Particle Irradiation (MOE), 
Institute of Frontier and Interdisciplinary Science, 
Shandong University, Qingdao, Shandong 266237, China}

\author{Michael Annan Lisa}
\affiliation{Department of physics, The Ohio State University, Columbus, Ohio 43210, USA}

\author{Xin-Nian Wang}
\affiliation{Key Laboratory of Quark and Lepton Physics (MOE) and Institute of
Particle Physics, Central China Normal University, Wuhan, 430079, China}
\affiliation{Nuclear Science Division, MS 70R0319, 
Lawrence Berkeley National Laboratory, Berkeley, California 94720}

\begin{abstract}
%
%
\end{abstract}


\maketitle

\noindent{\bf{Introduction: QGP as a perfect fluid}}

Recently, the global polarization of $\Lambda$ and $\bar\Lambda$ hyperons in heavy-ion collisions (HIC) 
has been observed~\cite{STAR:2017ckg} 
by the STAR Collaboration at the Relativistic Heavy Ion Collider (RHIC) 
in Brookhaven National Laboratory (BNL). 
The discovery confirms the theoretical prediction~\cite{Liang:2004ph} made more than ten years ago 
and indicates that  the quark gluon plasma (QGP) produced in HIC possesses a vorticity 
as high as $10^{21}$s$^{-1}$, much higher than any other fluid observed in nature. 
This opens a new window to study properties of QGP and a new direction in high energy heavy ion physics.

QGP is an extraordinary state of matter under strong interaction. 
As early as the 1970s, T.D. Lee and G.~C.~Wick suggested that~\cite{Lee:1974ma}
 ``by distributing high energy or high nucleon density over a relatively large volume", 
 one could temporarily restore broken symmetries of the physical vacuum
and possibly create a new state of dense nuclear matter. 
 At the same time, it was realized that 
 the asymptotic freedom in of quantum chromodynamics (QCD) implies the existence
of an ultra-dense form of matter where all known hadrons are expected to dissolve into 
 a plasma of their elementary constituents, the deconfined quarks and the gluons.


\begin{figure}[htbp]
\resizebox{3.3in}{1.4in}{\includegraphics{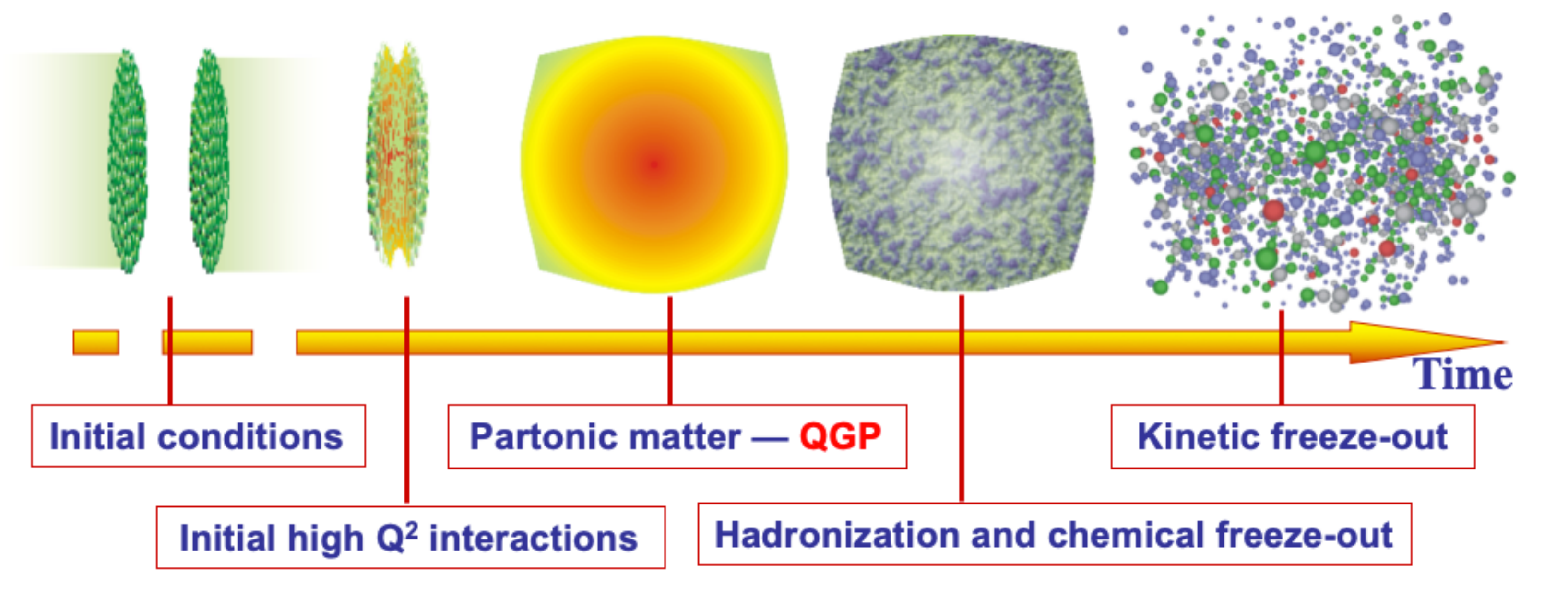}}
\caption{Illustration diagram of the evolution of a heavy ion collision.
The partonic matter system is found to be in the QGP state. }
\label{figHIC}
\end{figure}

The QGP state is thought to have existed in the early universe after the Big Bang~\cite{universe}. 
The transition from QGP to hadronic matter is believed to be 
one of several transitions occurring during the first few microseconds, 
when the temperature was of the order of hundreds of MeV ($\sim 10^{13}$K). 
It was pointed out that 
conditions required to create QGP 
could be achieved in the laboratory by colliding two heavy nuclei at high energies.  (See Fig.~\ref{figHIC}). 
Since then, QGP has been the focus of an extensive experimental program at SPS and LHC facilities at CERN,  
as well as RHIC at BNL~\cite{Baym:2001in}. 

The QGP observed in HIC has a number of remarkable properties manifested in 
azimuthal elliptical flow, jet quenching and other phenomena~\cite{Adams:2005dq}. 
It is also opaque to fast partons that carry color charge. 
These phenomena indicate that the QGP is a fluid with minimal viscosity, 
often called the ``perfect fluid"~\cite{Heinz:2013th}.

\begin{figure}[htbp]
\resizebox{2.45in}{1.8in}{\includegraphics{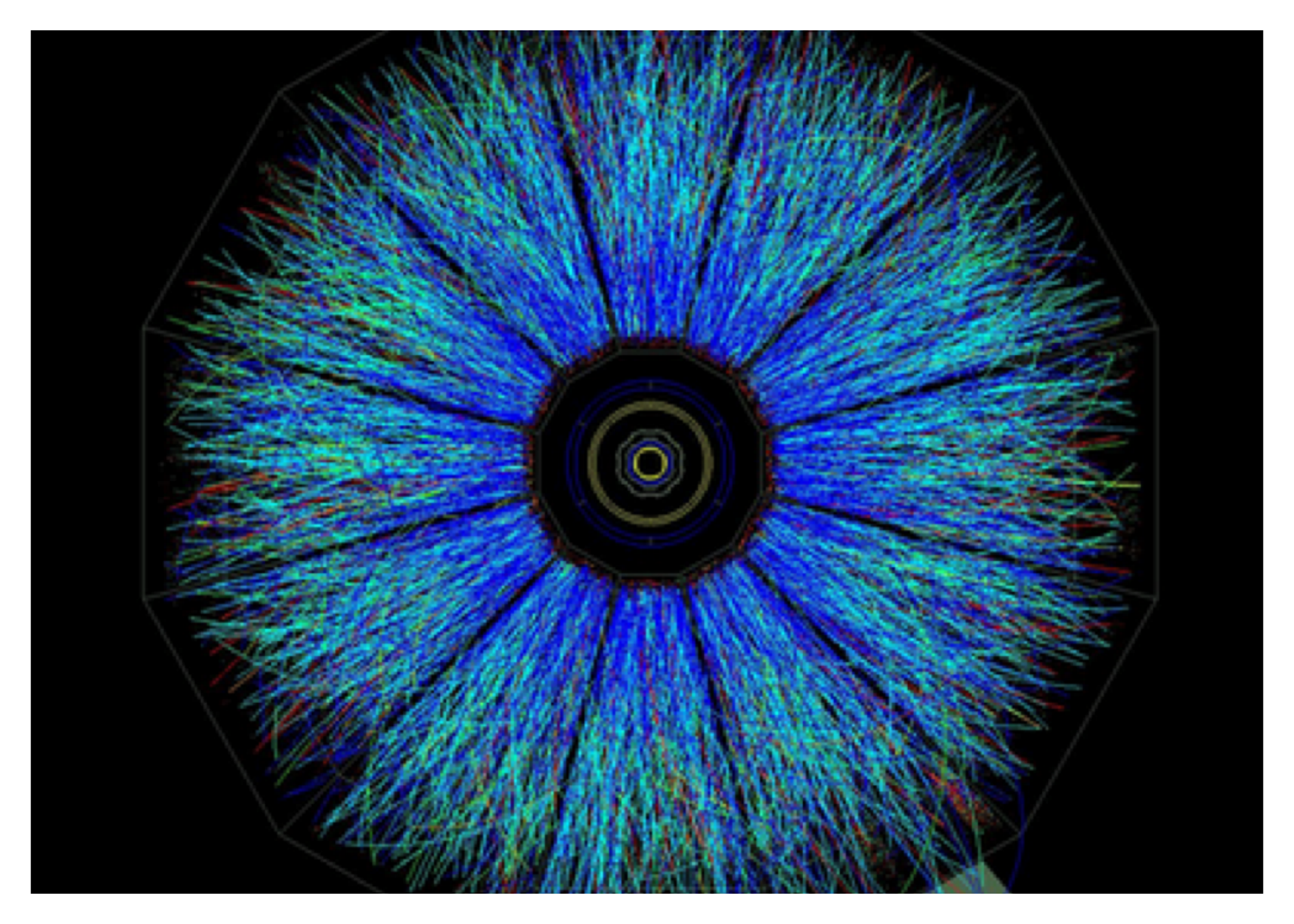}}
\caption{A heavy ion collision event at RHIC. 
In an average Au-Au collision at $\sqrt{s}=200$ AGeV, 
thousands hadrons are produced. 
Figure courtesy of the STAR collaboration: https://www.star.bnl.gov/.}
\label{figHICevent}
\end{figure}

As we discuss below, the study of polarization in HIC probes the fine rotational substructure 
of this perfect fluid. \\

\noindent{\bf{Global orbital angular momenta of QGP in HIC}}

Spin, as a fundamental degree of freedom of elementary particles, 
plays a very important role in modern physics and often bring us surprises.  
There are many well known examples in the field of high energy physics 
such as the so-called proton spin crisis~\cite{Aidala:2012mv} 
triggered by measurements on spin dependent structure functions in deeply inelastic lepton-nucleon scattering, 
the single transverse spin left-right asymmetry in inclusive hadron production in hadron-hadron collisions~\cite{SSA}, 
as well as the transverse hyperon polarization with respect to the production plane in unpolarized 
proton-proton and proton-nucleus collisions. 
Such studies are among the most active in strong interaction physics. 

The role of spin in HIC was largely ignored until it was realized~\cite{Liang:2004ph}
that the huge orbital angular momentum in these collisions may couple to 
spin degrees of freedom of the produced quarks.  
Exploring this phenomenon could provide an entirely new probe of the QGP, 
as well as studying the exchange of orbital and spin angular momentum under the most extreme conditions.

\begin{figure}[htbp]
\resizebox{3.9in}{2.1in}{\includegraphics{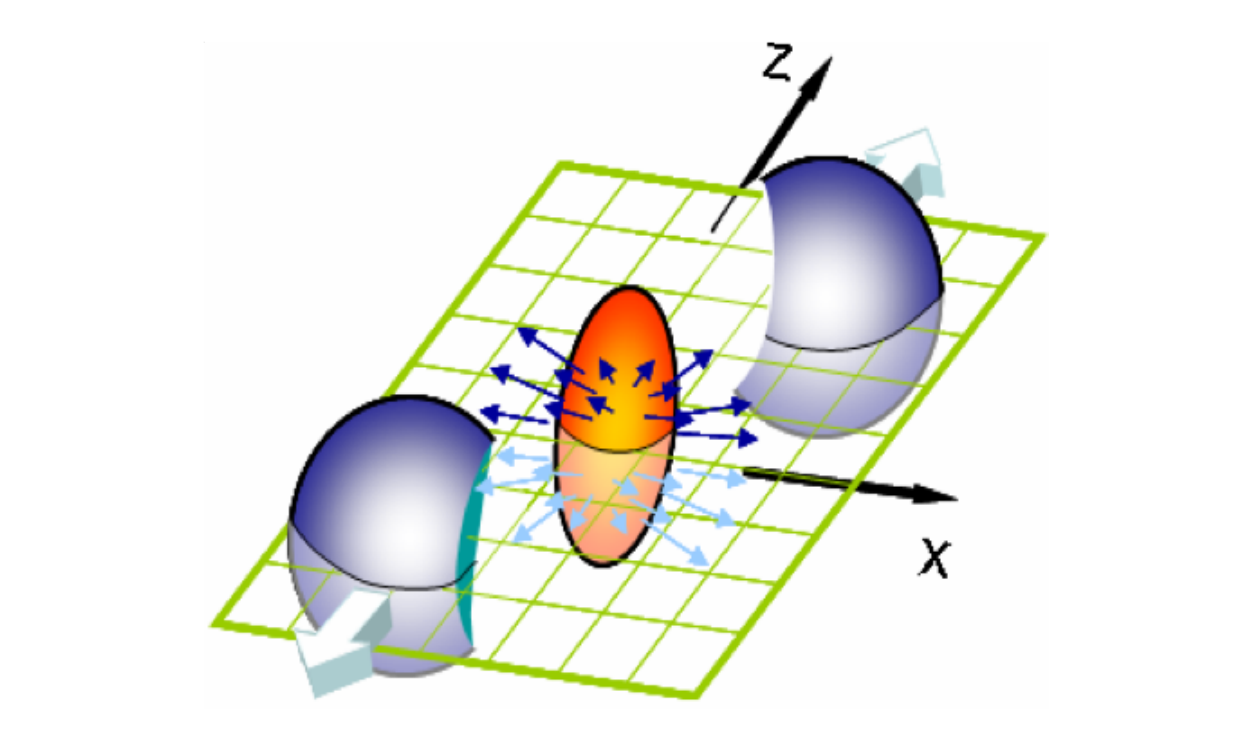}}
\caption{Illustration diagram for the reaction plane in a non-central heavy ion collision. 
In contrast to high energy $pp$ or $e^+e^-$ collisions, the reaction plane in 
a high energy heavy collision can be determined experimentally.}
\label{figreplane}
\end{figure}

\begin{figure}[htbp]
\resizebox{2.5in}{2.1in}{\includegraphics{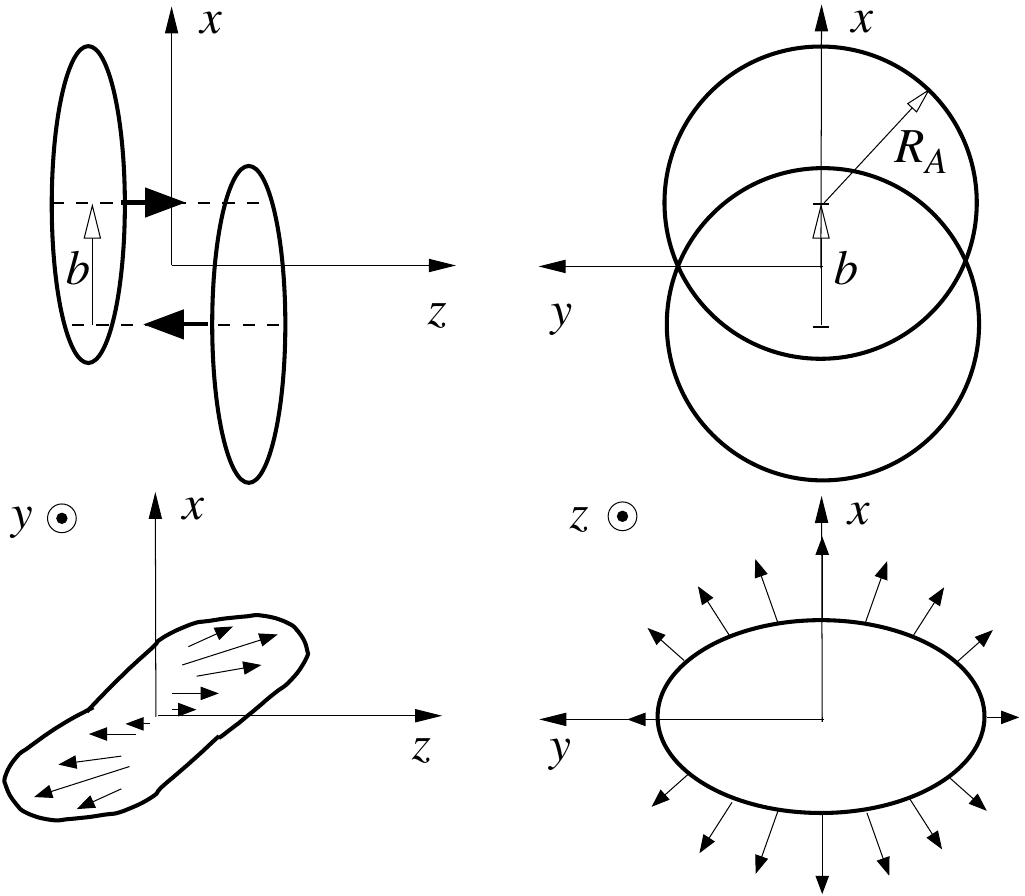}}
\caption{Illustration of the geometry and coordinate system 
of a non-central HIC with impact parameter $\mathbf{b}$. 
The global angular momentum of the produced matter is along the minus ${y}$ direction,
normal to the reaction plane. 
This figure is taken from \cite{Liang:2004ph}. }
\label{figgeo}
\end{figure}

The reaction plane and the geometry of a HIC as illustrated in Fig.~\ref{figreplane}
can be further clarified in Fig.~\ref{figgeo}. 
In the upper left panel, a nucleus (denoted the ``projectile") travels in the $+z$-direction, 
colliding with the ``target" traveling in the $-z$-direction. 
The impact parameter, $\mathbf{b}$ is the transverse component of the vector 
pointing from of the target to the projectile. 
We take $\mathbf{b}$ to define the $+{x}$-direction.
The normal of the reaction plane is given by 
$\mathbf{n}\propto {\mathbf p}\times{\mathbf{b}}$ is along the ${y}$-direction. 

\begin{figure}[htbp]
\resizebox{2.7in}{1.8in}{\includegraphics{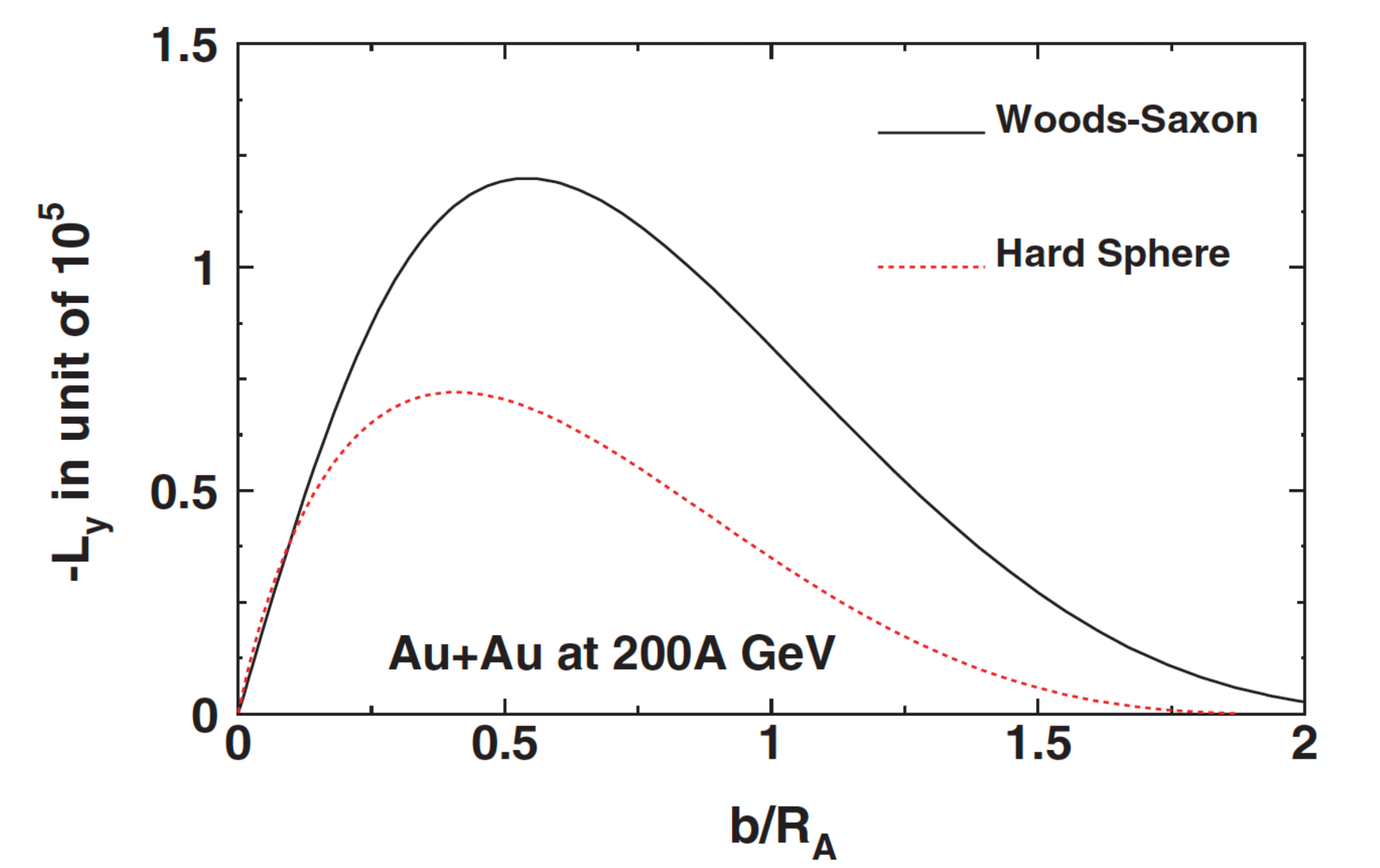}}
\caption{Global orbital angular momentum of the fireball formed in 
a non-central HIC as a function of the impact parameter. 
As in Fig.~\ref{figgeo}, $R_A=(1.12$ fm)$ A^{1/3}$ is the nuclear radius in a hard-sphere distribution.
This figure is taken from \cite{Gao:2007bc}.}
\label{figgly}
\end{figure}

The orbital angular momentum of the QGP, $L_y$, 
also known as the global angular momentum,  
in a non-central HIC has been estimated in~\cite{Liang:2004ph,Gao:2007bc} 
using a hard sphere or Wood-Saxon distribution of the nuclear matter in a nucleus. 
The results are shown in Fig.~\ref{figgly}. 
Here we see that the global orbital angular momentum of the QGP formed in HIC at RHIC 
can be as large as $10^5\hbar$.  
 
This global angular momentum of the partonic system can be manifested 
in the finite transverse gradient of 
the average longitudinal momentum $p_z$ per produced parton along the ${x}$-direction.
From the transverse gradient $dp_z/dx$, Ref.~\cite{Gao:2007bc} has also made 
an estimation of the so-called local orbital angular momentum 
for two partons in QGP with a transverse separation $\Delta x$. 
The  typical magnitude $l_0\sim -(\Delta x)^2 dp_0/dx$ 
in $Au+Au$ collisions at $\sqrt{s}=200$AGeV is 
$l_0\simeq -1.7$ for $\Delta x=1$fm and is indeed sizable. 

\begin{figure}[htbp]
\resizebox{3.0in}{2.0in}{\includegraphics{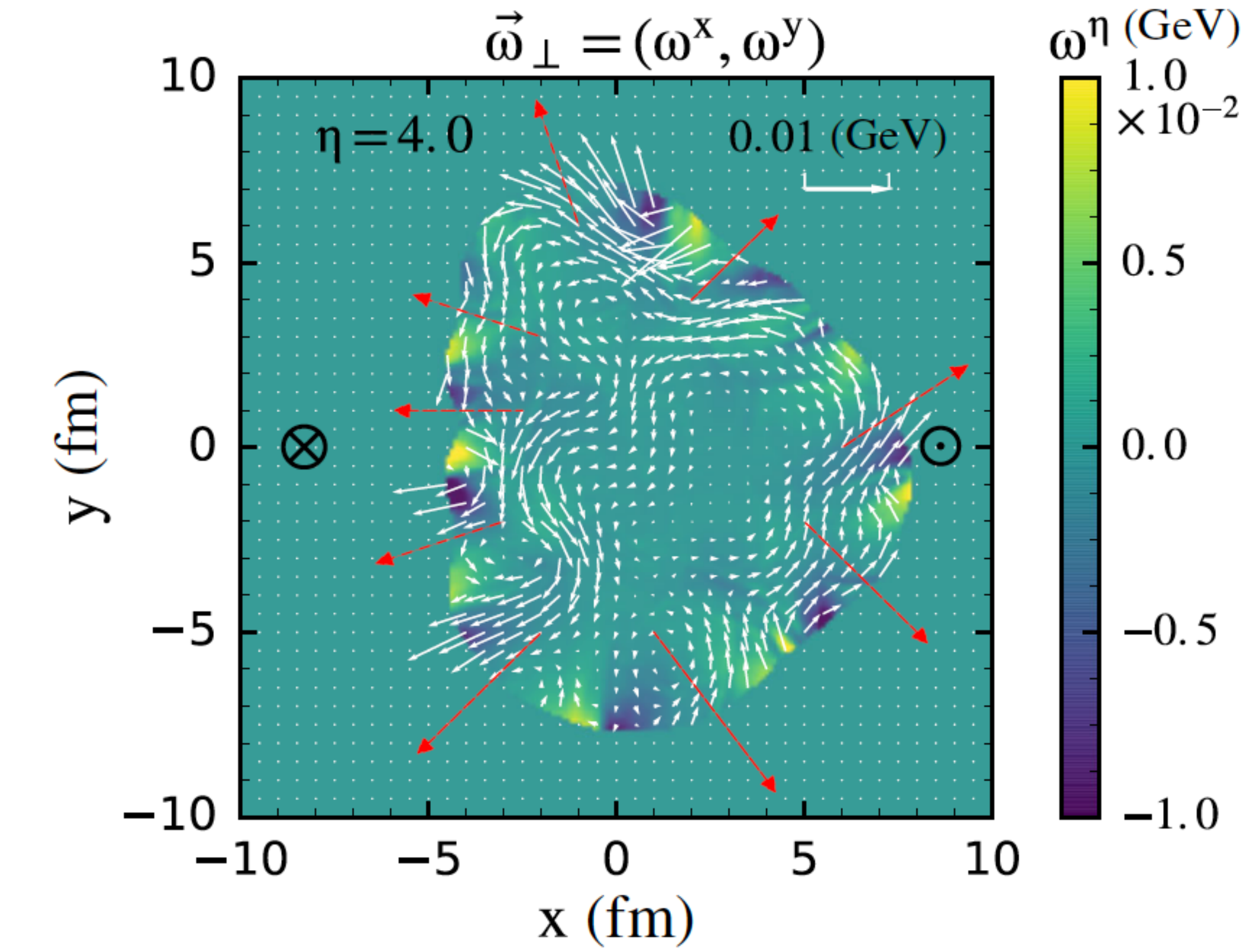}}
\caption{The local vorticity obtained in a hydrodynamical simulation. 
This figure is taken from~\cite{pang2016}.}
\label{figlocalvorticity}
\end{figure}

In more realistic hydrodynamic simulations of non-central collisions~\cite{pang2016}, 
local vorticities are generated through collected flow in a homogenous fluid as shown in Fig.~\ref{figlocalvorticity}. 
In the transverse plane of a non-central $Au$+$Au$ collision, the pattern of the local vorticity has a ring structure.
The net sum of the local vorticity is non-zero in the direction perpendicular to the reaction plane caused by 
the global angular momentum transferred to the fluid system.\\

\noindent{\bf{Spin-orbital coupling in a relativistic quantum system}}

It has been known for a long time that spin-orbital coupling is 
an intrinsic property of a relativistic fermionic quantum system. 
This can be seen most clearly from Dirac equation. 
Even for a free Dirac particle, the Hamiltonian, 
\begin{equation}
\hat H={\mathbf\alpha}\cdot\hat{\mathbf p}+\beta m,
\end{equation}
does not commute with the orbital angular momentum $\hat{\mathbf L}$ 
and the spin operator $\hat{\mathbf S}={\mathbf\Sigma}/2$ separately, but 
commutes with the total angular momentum $\hat{\mathbf J}=\hat{\mathbf L}+\hat{\mathbf S}$. 
This tells us clearly that the spin and the orbital angular momentum coupled to each other 
and should be able to transfer from one to another in a relativistic fermionic quantum system.
The strength of the spin-orbital coupling should be dependent on the strength of 
the interaction in the system considered. 
It should be different for a strongly interacting system from that for an electromagnetically interacting system.  

There are many well-known effects due to spin-orbital coupling in a system with electromagnetic interactions. 
The textbook example is the fine structure of atomic spectra. 
It plays also a very important role in modern spintronics in condensed matter physics where spin transport in 
the electromagnetically interacting system is studied. 

The transfer of spin polarization (magnetization) and orbital 
angular momentum (rotation) in electromagnetically interacting systems
was discovered just over a century ago. 
Barnet~\cite{barnett} found that rotation can cause magnetization at an object,  
while Einstein and deHaas showed that  a magnetic field may induce rotation. 
 
In systems with strong interactions, spin-orbital coupling also leads to many distinguished effects.  
One famous example is the nuclear shell model developed by Mayer and Jensen,  
where spin-orbit coupling plays a crucial role to produce the magic numbers of atomic nuclei.  
Currently, in the frontier of high energy spin physics, the orbital angular momentum is argued 
not only to contribute to the proton spin due to spin-orbital interactions but also   
can be responsible for the single-spin left-right asymmetries and other single-spin effects observed experimentally. 
The study of the role played by the orbital angular momentum is one of the core issues currently 
in high energy spin physics~\cite{Aidala:2012mv}.  \\

\noindent{\bf{Globally polarized QGP in HIC}}

It was realized that spin-orbit coupling in a strongly interacting QGP with large 
orbital angular momentum can lead to polarization of quarks and anti-quarks 
in non-central heavy-ion collisions. 

The theoretical calculations~\cite{Liang:2004ph} 
based on pQCD in the hot nuclear medium show that a quark can acquire 
a reasonably large polarization in the same direction 
as the initial orbital angular momentum after one collision with another parton in 
a fluid with non-vanishing local angular momentum or vorticity. 
This lead to the prediction that quarks and anti-quarks in 
QGP created in non-central HICs are polarized in the direction 
of the global angular momentum of the initial colliding system.  

Such a polarization~\cite{Liang:2004ph} is quite different from other cases in high energy physics
such as the longitudinal or the transverse polarization. 
The longitudinal polarization refers to the helicity or the polarization in the direction of the momentum; 
whereas the transverse polarization refers to directions perpendicular to the momentum, 
either in the production plane or along the normal of the production plane. 
These directions are all defined by the momentum of the individual particle.  
In contrast, the polarization discussed here refers to the normal of the reaction plane of an event. 
It is a fixed direction for one collision event and is independent of the direction of any particular hadron. 
Hence, in Ref.~\cite{Liang:2004ph}, this polarization was given a new name --- the global polarization, 
and the QGP was referred to the globally polarized QGP in non-central HIC. 

\begin{figure}[htbp]
\resizebox{3.1in}{2.1in}{\includegraphics{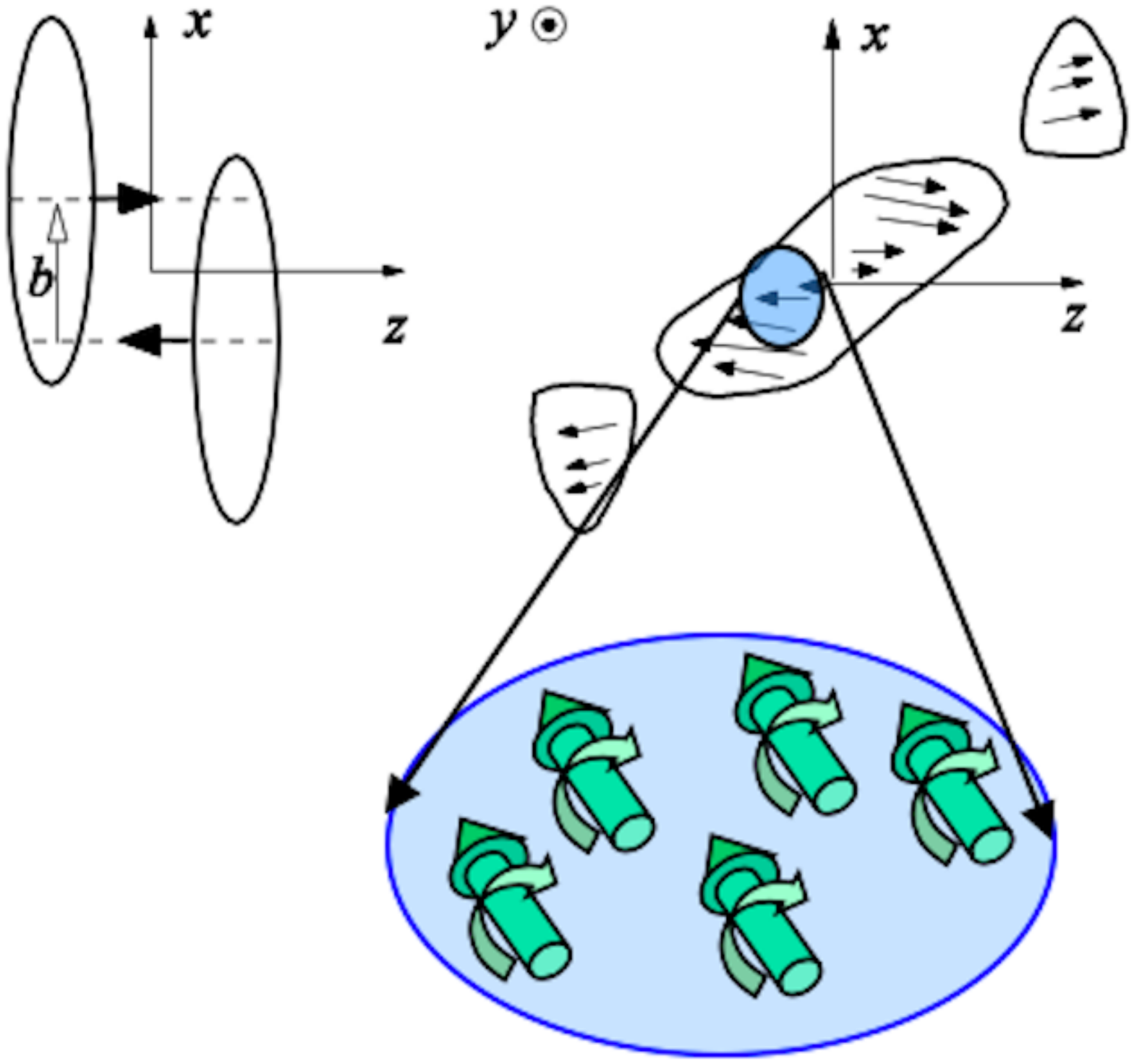}}
\caption{Illustration of the global polarization effect in non-central heavy ion collisions.}
\label{figGP}
\end{figure}

The global polarization of QGP can manifest itself in different observables. 
The most obvious one should be the global polarization of hadrons produced in the hadronization of the QGP.  
In Ref.~\cite{Liang:2004ph}, the global polarization of produced hyperons are estimated. 
The spin alignment of vector mesons (hadrons with spin-1) has also been calculated~\cite{Liang:2004xn}. 
Many other observables and calculations based on different approaches 
have also been discussed in the literature. 
See e.g. a summary by Qun Wang in Quark Matter 2017~\cite{Wang:2017jpl}. \\

\noindent{\bf{STAR experiments and the vorticity of QGP}}

These novel predictions on the global polarization attracted immediate attention, 
both experimentally and theoretically. 
A new preprint~\cite{Voloshin:2004ha} only three days after the first prediction~\cite{Liang:2004ph} 
attempted to extend the idea to other reactions. 
Experimentalists in the STAR collaboration had also started measurements shortly afterward. 

Both hyperon polarization and vector meson spin alignment can be measured 
via the angular distribution of decay products.
For hyperon polarization, the polarization is measured by the so-called self-spin analyzing weak decay. 
For example, for $\Lambda$ hyperon, 
this can be measured by studying the weak decay $\Lambda\to p\pi^-$, 
\begin{equation}
\frac{dN}{d\Omega}=\frac{N}{4\pi}(1+\alpha P_\Lambda\cos\theta),
\end{equation}
where $\theta$ is the decay angle of the proton with respect to 
the polarization direction of $\Lambda$ in its rest frame,  
$\alpha$ is a known constant that is called a decay parameter, 
and $P_\Lambda$ is the magnitude of the polarization. 

Measurements of global $\Lambda$ hyperon polarization 
and $K^*$ and $\phi$ vector meson spin alignments of $K^*$ and $\phi$   
attempted soon after the publication of the theoretical predictions~\cite{Liang:2004ph,Liang:2004xn}. 
Studies of both aspects have advantages and disadvantages. 
Hyperon polarization is a linear effect where 
the polarization for directly produced $\Lambda$ is equal to that of quarks.
The spin alignment of vector meson is a quadratic effect proportional to the square of quark polarization. 
Hence the magnitude of effects of vector meson spin alignment should be much smaller than that of hyperon polarization.  
However, to measure the polarization of hyperon, one has to determine the direction of  
the normal of the reaction plane, which is not needed for measurements of vector meson spin alignments. 
Also the feed down effects due to decay contributions to vector mesons are negligible but not for $\Lambda$ hyperons. 
 
Although there were some promising indications, the results obtained 
in the early measurements~\cite{Abelev:2007zk,Abelev:2008ag} by the STAR collaboration 
both on $\Lambda$ hyperon polarizations and vector meson spin alignments were  
consistent with zero within the large errors. 
STAR measurements continued during the beam energy scan (BES) experiments   
and positive results were obtained with improved accuracies~\cite{STAR:2017ckg}. 

\begin{figure}[htbp]
\resizebox{3.1in}{1.8in}{\includegraphics{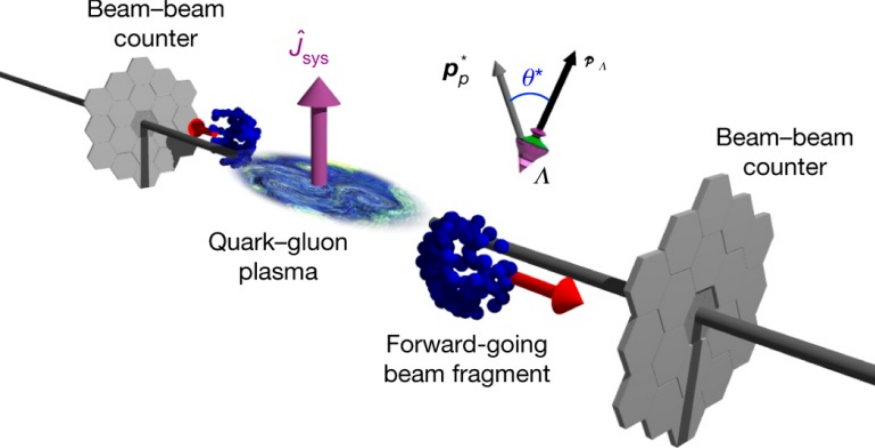}}
\caption{A sketch of a heavy ion collision in the STAR detector system.
The reaction plane was determined by measuring 
the sidewards deflection of the forward and backward going fragments 
and particles in the beam-beam counter detectors.
This figure is taken from~\cite{STAR:2017ckg}.}
\label{figStarlayout}
\end{figure}

A sketch of a HIC in the STAR detector system 
is given in Fig.~\ref{figStarlayout}, 
where the reaction plane was determined by measuring 
the sidewards deflection of the forward and backward going fragments 
and particles in the beam-beam counter detectors. 

\begin{figure}[htbp]
\resizebox{2.56in}{2.2in}{\includegraphics{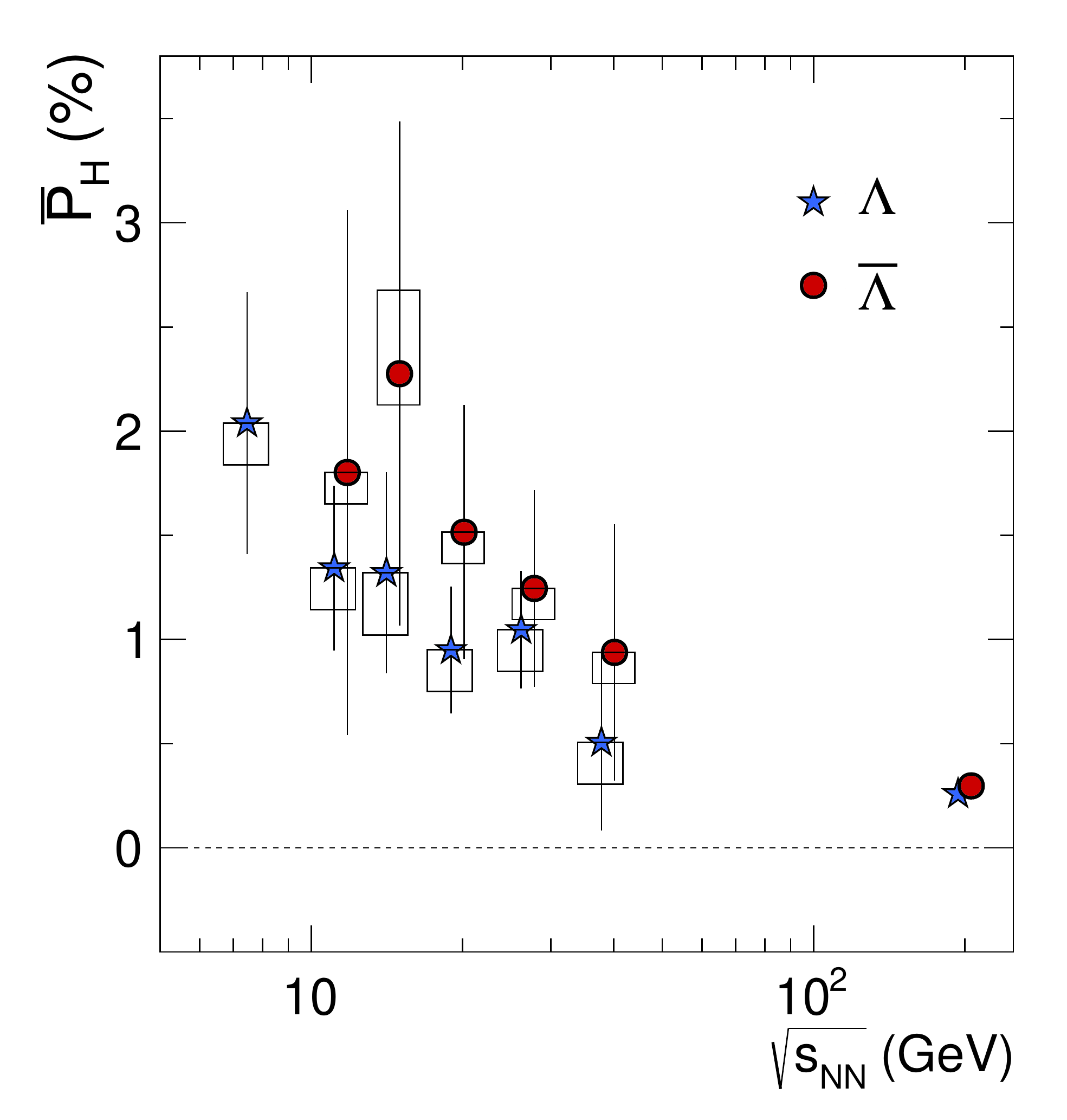}}
\caption{Global polarizations for $\Lambda$'s (blue stars) and $\bar\Lambda$'s (red circles)
obtained by the STAR Collaboration. Figure adopted from~\cite{STAR:2017ckg}. }
\label{figstardata}
\end{figure}

Fig.~\ref{figstardata} show the results obtained by the STAR Collaboration for 
the global polarization of $\Lambda$ and $\bar\Lambda$ as functions of the collision energy.  
The obtained value averaged over energy is $1.08 \pm 0.15  (stat) \pm 0.11  (sys)$  per cent 
and  $1.38 \pm 0.30 (stat) \pm  0.13 (sys)$ per cent for $\Lambda $ and $\bar\Lambda$, respectively.

\begin{figure}[htbp]
\resizebox{3.3in}{1.5in}{\includegraphics{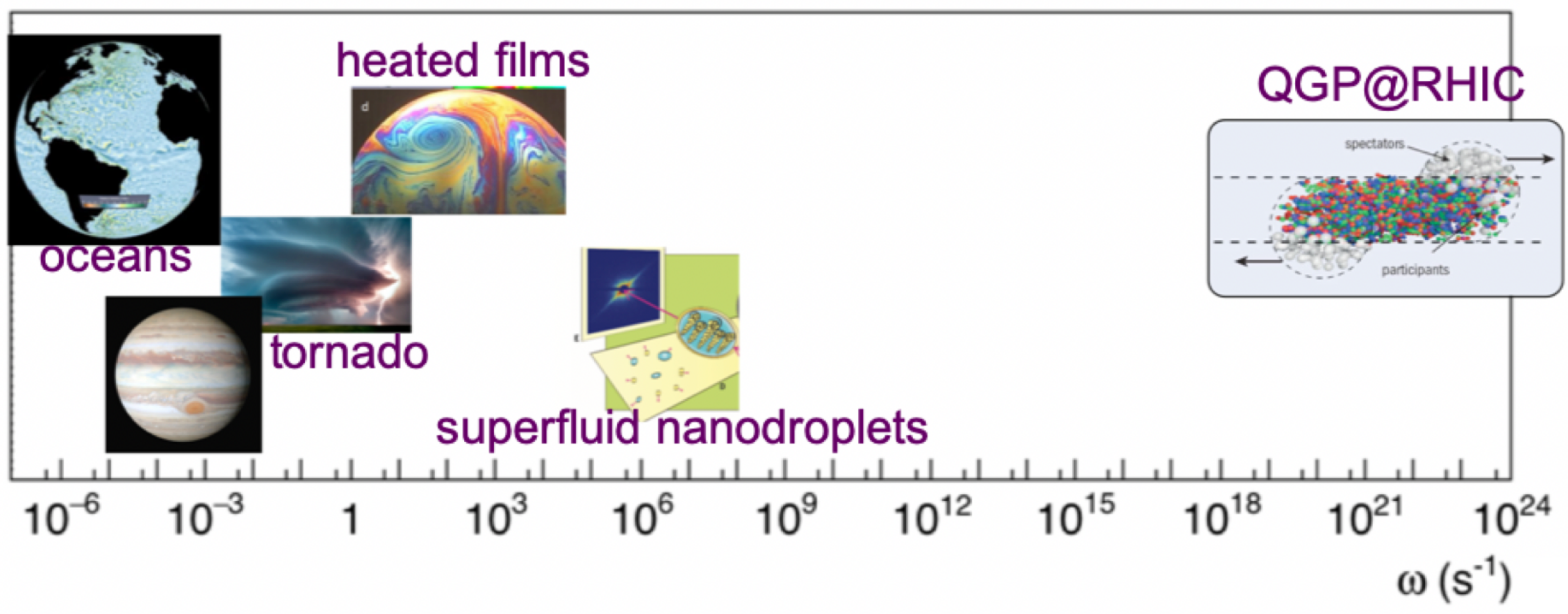}}
\caption{Illustration of scales of vorticities of different fluids observed in nature. }
\label{figvorticity}
\end{figure}

If QGP is in equilibrium and we can treat it as a vortical ideal fluid consisting of quarks and anti-quarks, 
the global polarization of $\Lambda$ hyperon is directly related to the vorticity of the system~\cite{Becattini:2016gvu}.
The fluid vorticity may be estimated from the data using the relation given in 
the hydro-dynamic model, 
and it leads to a vorticity $\omega\approx(9\pm1)\times 10^{21} s^{-1}$. 
This far surpasses the vorticity of all other known fluids
(See Fig.~\ref{figvorticity}.).  
It was therefore concluded that QGP created in HIC is the most vortical fluid in nature observed yet.\\

\noindent{\bf{The outlook}}

The early theoretical prediction~\cite{Liang:2004ph} 
and discovery of global spin polarization in non-central HIC by STAR Collaboration~\cite{STAR:2017ckg}  
open a new window to study properties of QGP and a new direction in high energy heavy ion physics. 
With much higher statistics, the STAR Collaboration has repeated 
measurements~\cite{Adam:2018ivw} in Au-Au collisions at 200AGeV and obtained  positive results 
with much higher accuracies.
The ALICE Collaboration at the Large Hadron Collider (LHC) has also carried out similar measurements 
in Pb-Pb collisions~\cite{Acharya:2019ryw}.
Other efforts on measurements of vector meson spin alignments have also been attempted.
The STAR Collaboration has just finished major detector upgrades and started 
the beam energy scan experiments at phase II (BES II). 
The successful detector upgrade with improved 
inner time projection chamber (iTPC) and event plane detector (EPD) 
will be crucial to the measurements of global hadron polarizations. 
The STAR BES II will provide an excellent opportunity to study the global polarization in HIC and 
we expect new results with higher accuracies in next years.  

There are also many exciting studies in this connection such as different approaches to calculate 
the global polarizations, other measurable effects on global polarizations, and also other effects 
in connection with the huge orbital angular momenta of the colliding systems in HIC. 
Interested readers are referred to recent overviews such as~\cite{Wang:2017jpl}.

\begin{figure}[tbtp]
\resizebox{3.3in}{1.5in}{\includegraphics{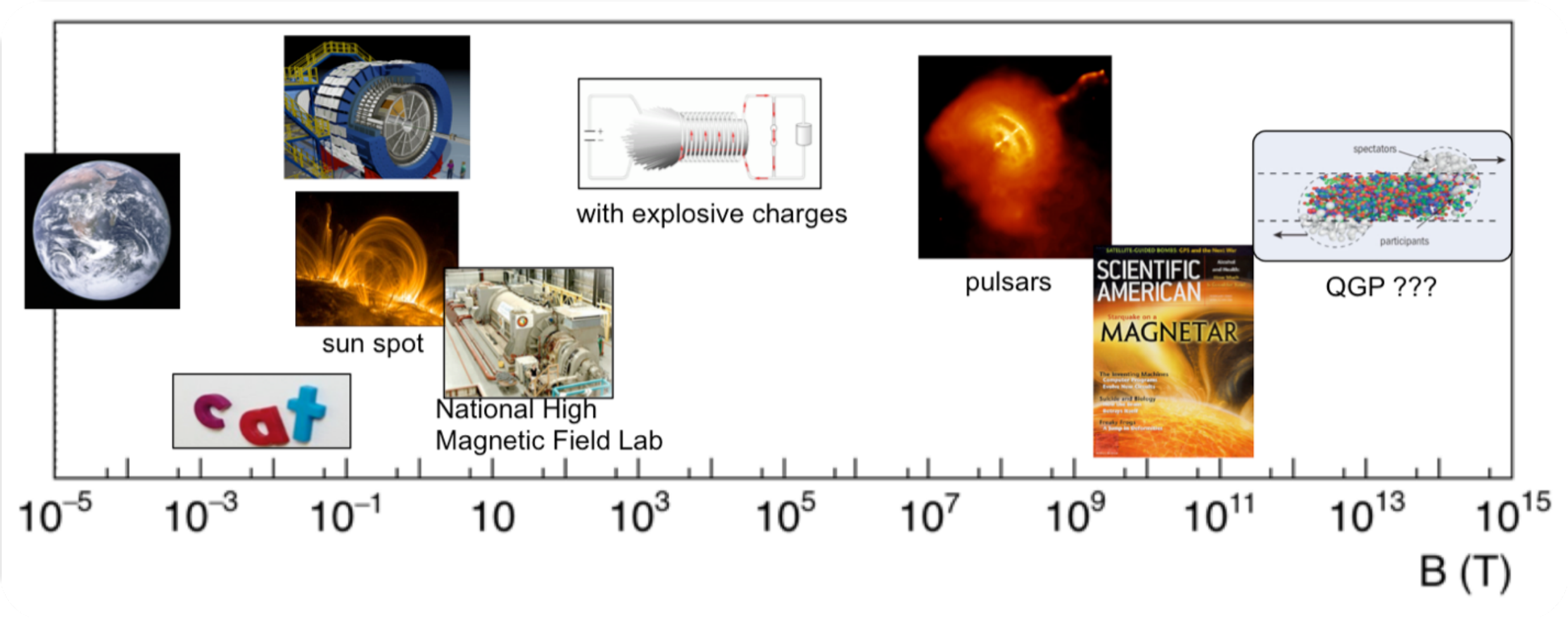}}
\caption{Illustration of scales of magnetic fields of different systems in nature.}
\label{figmag}
\end{figure}

In addition to vorticity and rotation, these polarization studies may provide 
a unique probe of the huge magnetic fields that are expected in HIC.  
The relativistic charged nuclei should produce fields on order $10^{14}-10^{16}$~Tesla, at least instantaneously.  
The superconducting nature of the QGP may extend the lifetime of this field as the system evolves.  
While vortical or spin-orbit coupling tends to align all spins in the same direction, 
magnetic coupling can produce a ``fine structure" to the coupling systematics~\cite{Becattini:2016gvu}.  
In particular, a strong magnetic field would enhance the polarization of $\overline{\Lambda}$ 
and slightly suppress that of $\Lambda$.  
Indeed, the data in Fig.~\ref{} suggests just such a fine-structure pattern, 
and if errorbars are ignored, would indicate $B\sim10^{14}$~T, 
higher than any other magnetic field in the universe (see Fig.~\ref{figmag}).  
However, much smaller uncertainties-- available with the new BES-II data-- will be needed to resolve the issue. \\ 

\noindent{\bf{Acknowledgements}}:
We thank in particular the STAR Collaboration for providing the excellent experimental environments 
and for the standing efforts in carrying the measurements in this direction. 
This work was supported in part by 
the National Natural Science Foundation of China  (Nos. 11890713, 11675092, 11861131009 and 11890714), 
by the U.S. National Science Foundation award 1614835 and DOE under Contract No. DE-AC02- 05CH11231.

\end{document}